\documentclass[12pt]{article}
\usepackage{amssymb,amsmath,graphicx}
\title{Baryon Asymmetry, Inflation and Squeezed States.}
\author{Bindu A.Bambah,K.V.S.Shiv Chaitanya,\\ School of Physics,
University of Hyderabad,\\Hyderabad 500
046,India\\and\\C.Mukku,\\Indian Institute of Information
Technology, \\Gachibowli, Hyderabad-500 046,India}
\begin{document}
\maketitle
\begin{abstract}
We use the general formalism of squeezed rotated states to
calculate baryon asymmetry in the wake of inflation through
parametric amplification. We base our analysis on a B and CP
violating Lagrangian in an isotropically expanding universe.
The B and CP violating terms originate from the coupling of
complex fields with non-zero baryon number to a complex
background inflaton field. We show that a differential
amplification of particle and anti-particle modes gives rise
to baryon asymmetry.
\end{abstract}
\section{Introduction}

Baryon asymmetry is one of the challenging problems in the
standard model of cosmology \cite{Weinberg,kolb} . It is
characterized by the ratio $\eta=\frac{n_b}{n_{\gamma}} $, where
$n_b$ is the number of baryons and $n_{\gamma}$ is the number of
photons in the universe. The present value of the asymmetry in the
universe is $\eta \simeq 10^{-10}$. Three conditions postulated by
Sakharov \cite{sakharov} are sufficient to guarantee baryon
asymmetry. These are  (i) Baryon number(B) violation (ii) charge
and parity (CP) violation (iii) and the presence of nonequilibrium
processes. These conditions are satisfied at very high
temperatures, i.e. at Grand Unified Theory (GUT) scales. Thus the
first theory of baryogenesis proposed was that of GUT baryogenesis
\cite{yoshimura, weinberg2,dolgov}. In GUT baryogenesis, baryon
asymmetry is generated through the decay of baryon number
violating bosons known generically as X bosons. The general
feature of GUT baryogenesis is that the characteristic energy
scale for the processes with B-nonconservation is extremely high,
close to the Planck scale, $M_{GUT} = 10^{16 }GeV$, thus, it would
be kinematically difficult to produce these Bosons in a thermal
environment. At temperatures higher than the GUT scale, the rate
of their production would be smaller than of the expansion of the
universe  and any generated baryon asymmetry is presumed to be
wiped out by this  expansion. If the  temperature of the universe
could always be smaller than the GUT scale, the corresponding,
thermally produced X-bosons would never have been abundant and
their role in baryogenesis would be negligible. Furthermore,
together with B violating interactions that produce asymmetry in
GUTS, there should be proton decay and neutron antineutron
oscillations, which have, unfortunately, not been detected
experimentally. Since then, many other scenarios of baryogenesis
have been developed with baryon non-conservation at much lower
energies. One such theory is electroweak baryogenesis where
essential physical processes take place at an energy scale of
around 100 GeV. The problem with electroweak baryogenesis is that
the CP violation that generates the asymmetry is at least 12
orders in magnitude smaller than the observed asymmetry. One
proposed solution to overcome these difficulties with baryogenesis
models is to consider baryon asymmetry in the context of
inflation. It is  believed that an inflationary phase precedes the
hot phase of cosmological evolution. In the inflationary phase,
the energy density is dominated by the contribution from scalar
field(s) called inflaton(s) \cite{inflation}. A phase transition
from an inflationary phase to a thermal phase may provide, in
principle, the large temperatures needed for baryogenesis. During
inflation, all matter and radiation are inflated away. Thus when
inflation is over, there must be a reheating process, when the
energy stored in the inflaton(s) is converted into thermal
radiation. There are several possibilities for this reheating
process. The inflaton may decay perturbatively into light
particles, which thermalize eventually. There are also
non-perturbative models of reheating known as "preheating "
models. These are of two types. In one type of model the process
of parametric resonance \cite{brand1, brand} may greatly enhance
the production of particles  \cite{kofman}. In another, a rapid
quench in the motion of the inflaton may lead to a spinodal
instability and results in fast tachyonic preheating
\cite{felder}. The preheating process is a far from equilibrium
process forcing the universe into a non-equilibrium state after
the end of inflation and before thermalization takes place. Thus
it fulfills the third criterion for baryogenesis. It has been
shown that some baryon-number violating processes similar to
sphaleron transitions take place at preheating
\cite{kusenko,garcia}, as well as tachyonic preheating
\cite{smit}. An interesting feature of the preheating scenario is
that the production of particles with masses greater than the
inflaton mass is possible. In particular, the non-thermally
produced particles could include the baryon number violating gauge
and Higgs bosons of grand unified theories (GUTs) or right-handed
neutrinos, resurrecting the possibility of GUT-scale baryogenesis
or leptogenesis .

Preceding all of this, Papastamatiou and Parker \cite{pp} proposed
a simple model in which the matter-antimatter asymmetry  created
by a baryon and CP violating Lagrangian in an isotropically
expanding universe is found by a perturbative expansion. Taking an
initial state in which no particles participating in the
asymmetric interaction are present, they found that
matter-antimatter asymmetry is produced during a stage when the
radius of the universe is small with respect to its present value.
In this work, we have adapted their work to consider the asymmetry
that might be created in a more realistic universe that is
initially inflating and then enters a rapid preheating phase
either by parametric resonance or tachyonic preheating.
Furthermore we use a more realistic source of CP violation,
namely, a complex time varying background field which we can
identify with a classical inflaton field , like in ref.
\cite{ranga}. Preheating is nothing but particle creation in the
oscillating background, or one with spinodal instability
(tachyonic preheating). Although particle creation or annihilation
generally occurs in a time-dependent background, the number of
created particles grows exponentially either when the background
is periodic in time or when the coupling constant changes
suddenly. This results in far from equilibrium particle production
and there is an amplification of long wavelength modes.
 Therefore, if we add the ingredients of CP and B violation, there are chances for generating
the matter-antimatter asymmetry of the universe in these eras.
Recently, Ref. \cite{ranga,japanese} appeared in which the authors
discuss the generation of the baryon asymmetry during reheating.
In one of these \cite{japanese} baryon asymmetry was generated by
considering particle creation in the interaction of scalar fields
with an  oscillating inflaton  background. In both these papers,
baryon asymmetry was evaluated perturbatively, and the
perturbation theory limited the produced asymmetry. In this paper,
 the first of a series, we extend the work in
\cite{ranga,japanese} to include both expansion and
non-perturbative effects  and look for baryon asymmetry in the
wake of inflation through parametric amplification. We calculate
the effect of CP violation without resorting to perturbation
theory. This allows us to compute non-perturbative out of
equilibrium generation of baryon asymmetry. We do so by
considering  the Lagrangian of Papastamatiou and Parker in an
expanding inflationary universe and treat particle production in
the general formalism of squeezed rotated states.

\section{The Formalism}

The most general Lagrangian density with a baryon non-conserving
interaction in an expanding inflationary cosmology has been given
in  \cite{pp,ranga} as \begin{eqnarray}
 L&=&\sqrt{-g}[g^{\mu\nu}\partial_\mu\phi^*\partial_\nu\phi+
 g^{\mu\nu}\partial_\mu\psi^*\partial_\nu\psi-
 (m_{\phi}^2+\xi_{\phi}R)\phi^*\phi-
(m_{\psi}^2+\xi_{\psi}R)\psi^*\psi- \nonumber \\
&&+g^{\mu\nu}\partial_\mu\eta^*\partial_\nu\eta
 -(m_{\eta}^2+\xi_{\eta}R)\eta^*\eta-V(\eta)
-\lambda(\eta^2\phi^*\psi+\eta^{*2}\psi^*\phi)],
\end{eqnarray}
 where, $m_{\phi,\psi}$ are
masses of fields $\phi$ and $\psi$. $\phi$ is a complex scalar
field that carries a baryon number 1 and $\psi$ is a complex
scalar field that carries a baryon number 0 (which can be treated
as an antilepton field). The field $\eta$ is a  minimally coupled
 complex inflaton field, R is the Ricci scalar and $\xi$ is
curvature coupling ($\xi = {1\over 6}$ gives the traditional
conformal coupling of scalar fields to gravity). The sole purpose
of the inflaton field $\eta$ is to provide for inflation and
therefore, it is treated as a background classical field while
quantum excitations are considered for the other scalar fields.
Our prime interest in this paper is to explore the effects of the
inflaton field and the expansion of the Universe on the baryon
assymetry. To this end, we consider the background gravitational
field to be that of a standard FRW cosmology with an expansion
parameter $a(t)$. This Lagrangian can be thought of as an
effective Lagrangian which contains terms that violate baryon
number by means of a complex field which carries a baryon number
interacting with the background inflaton field. This results in a
net asymmetry between the produced particles and antiparticles
represented by these complex fields, which gives rise to a net
baryon asymmetry \cite{pp}. It is this particle -antiparticle
asymmetry that we now proceed to calculate.

 In order to see the salient features of
baryogenesis, we simplify the Lagrangian density through the
following assumptions. The inflaton field  $\eta$  is a classical
homogeneous time dependent complex background field, whose
temporal evolution is determined by a Euler Lagrange equation of
motion, with the inflaton potential $V(\eta)$ in the
Friedman-Robertson-Walker metric $ds^2=dt^2-a^2(t)dx^2,$  given by
   \begin{equation}
\ddot{\eta}+3\frac{\dot{a}}{a}\dot{\eta}+\frac{\partial
V}{\partial\eta} = 0.
\end{equation}
 For this paper, we consider the minimal coupling to the background
 gravitational field ($\xi_\phi = \xi_\psi = \xi_\eta = 0$).
 The action is given  by
 \begin{eqnarray}S&=&
 \int d^3x dta^3(t)({1\over 2}(\dot{\phi}^*)(\dot{\phi})-
{1\over 2a^2(t)}(\nabla\phi^*)({\nabla\phi})+{1\over
2}(\dot{\psi}^*)(\dot{\psi})-{1\over
2a^2(t)}(\nabla\psi^*)(\nabla\psi)\nonumber \\&-&
 m_\phi\phi^*\phi-
m_\psi\psi^*\psi  - \lambda(\eta^2\phi^*\psi+\eta^{*2}\psi^*\phi)).
\end{eqnarray}
Applying a Legendre transformation we can write the Hamiltonian as:
\begin{eqnarray}
H &=& \int d^3x dt  a^3[{1\over a^6} \pi^*_\phi \pi_\phi +
{1\over a^2} \nabla\phi^*\nabla\phi + {1\over a^6}
\pi^*_\psi \pi_\psi + {1\over
a^2}\nabla\psi^*\nabla\psi + m^2_\phi\phi^*\phi
+m^2_\psi\psi^*\psi \nonumber \\&+&
\lambda(\eta^2\phi^*\psi+\eta^{*2}\psi^*\phi)],
\end{eqnarray}
where, $\pi_\phi$ and
$\pi_\psi $ are the canonical
momenta of the $\phi$ and $\psi$ fields.  We canonically quantize
this Hamiltonian. Since we are treating the inflaton field as a
classical complex background field we quantize only the $\phi$ and
$\psi$ field by  introducing the mode expansions
\begin{eqnarray}
\psi(x^\mu) &=& \int d\tilde{k}[a_k^\psi e^{-ik\cdot x} +
b^{\dagger\psi}_{k} e^{ik\cdot x}],\\ \psi^*(x^\mu) &=& \int
d\tilde{k}[a^{\dagger\psi}_{k} e^{ik\cdot x} + b_k^\psi
e^{-ik\cdot x}],\\ \phi(x^\mu) &=& \int d\tilde{k}[a_k^\phi e^{-ik\cdot x}
 + b^{\dagger\phi}_{k} e^{ik\cdot x}],\\ \phi^*(x^\mu)
&=& \int d\tilde{k}[a^{\dagger\phi}_{k} e^{ik\cdot x} +
b_k^\phi e^{-ik\cdot x}],
\end{eqnarray}
here
\begin{eqnarray}
k\cdot x=k_\mu x^\mu =\omega t -k_ix_i,\\ d\tilde{k} = {d^3kdt\over
[(2\pi)^3 2\omega_k]^{1\over 2}},
\end{eqnarray}
and it is clear from the line element given above, the signature of the metric is (+,-,-,-).\\
The quantized Hamiltonian is given as
\begin{eqnarray}
H&=&\int d^3k [{\omega_{\psi}\over a^3} a^{\dagger\psi}_ka^\psi_k
+{\omega_{\phi}\over a^3}
a^{\dagger\phi}_ka^\phi_k+{\lambda\eta^2a^3\over
2\sqrt{\omega_\phi\omega_\psi}}
a^{\dagger\psi}_ka^\phi_k+{\lambda\eta^{*2}a^3\over
2\sqrt{\omega_\phi\omega_\psi}}
a^{\dagger\phi}_ka^\psi_k]\nonumber\\ &&+[{\omega_{\psi}\over
a^3}b^\psi_{-k} b^{\dagger\psi}_{-k} +{\omega_{\phi}\over a^3}b^\phi_{-k}
b^{\dagger\phi}_{-k}+{\lambda\eta^2a^3\over
2\sqrt{\omega_\phi\omega_\psi}}
b^\phi_{-k}b^{\dagger\psi}_{-k}+{\lambda\eta^{*2}a^3\over
2\sqrt{\omega_\phi\omega_\psi}}
b^\psi_{-k}b^{\dagger\phi}_{-k}]\nonumber\\ &&+{\lambda\eta^2a^3\over
2\sqrt{\omega_\phi\omega_\psi}} [a^{\dagger\psi}_k
b^{\dagger\phi}_{-k}+b^{\psi}_k a^{\phi}_{-k}]
+{\lambda\eta^{*2}a^3\over 2\sqrt{\omega_\phi\omega_\psi}}
[a^{\dagger\phi}_k b^{\dagger\psi}_{-k}+b^{\phi}_k a^{\psi}_{-k}],
\end{eqnarray}
where
\begin{eqnarray}
\frac{\omega_\phi^2}{a^6} &=& \underline{k}^2 + m_\phi^2, \\
 \frac{\omega_\psi^2}{a^6} &=& \underline{k}^2 + m_\psi^2,
\end{eqnarray}
here $\underline{k} = \frac{k}{a}$ is the physical wave (co-wave) number of the mode, k is the comoving wave number.\\
We define the following
\begin{equation}
\frac{\Omega_\psi^2(t)}{a^6} = \frac{\omega_\psi^2}{a^6} +
\sqrt{\frac{\omega_\psi}{\omega_\phi}}\lambda\eta^2(t),
\end{equation}

\begin{equation}
\frac{\Omega_\phi^2(t)}{a^6} = \frac{\omega_\phi^2}{a^6} +
\sqrt{\frac{\omega_\phi}{\omega_\psi}}\lambda\eta^{*2}(t),
\end{equation}
and write the Hamiltonian as
\begin{eqnarray}
H&=&\int d^3k [{\omega_{\psi}\over a^3} a^{\dagger\psi}_ka^\psi_k
+{\omega_{\phi}\over a^3}
a^{\dagger\phi}_ka^\phi_k+\frac{\omega_\psi}{a^3}
 \left[\frac{\Omega_\psi^2(t)}{\omega_\psi^2} - 1\right]
a^{\dagger\psi}_ka^\phi_k+\frac{\omega_\phi}{a^3}
\left[\frac{\Omega_\phi^2(t)}{\omega_\phi^2} - 1\right]
a^{\dagger\phi}_ka^\psi_k]\nonumber\\ &&+[{\omega_{\psi}\over
a^3}b^\psi_{k} b^{\dagger\psi}_{k} +{\omega_{\phi}\over
a^3}b^\phi_{k} b^{\dagger\phi}_{k}+\frac{\omega_\psi}{a^3}
\left[\frac{\Omega_\psi^2(t)}{\omega_\psi^2} - 1\right]
b^\phi_{k}b^{\dagger\psi}_{k}+\frac{\omega_\phi}{a^3}
\left[\frac{\Omega_\phi^2(t)}{\omega_\phi^2} - 1\right]
b^\psi_{k}b^{\dagger\phi}_{k}]\nonumber\\
&&+\frac{\omega_\psi}{a^3}
\left[\frac{\Omega_\psi^2(t)}{\omega_\psi^2} - 1\right]
[a^{\dagger\psi}_k b^{\dagger\phi}_{-k}+b^{\psi}_{-k} a^{\phi}_{k}]
+\frac{\omega_\phi}{a^3} \left[\frac{\Omega_\phi^2(t)}{\omega_\phi^2}
- 1\right] [a^{\dagger\phi}_k b^{\dagger\psi}_{-k}+b^{\phi}_{-k}
a^{\psi}_{k}].
\end{eqnarray}

To see the symmetries of this Hamiltonian, which
will aid us in its diagonalization, we define the operators:
\begin{eqnarray} N_{1} =a^{\dagger\psi}_ka^\psi_k,\; \;
N_{2}=a^{\dagger\phi}_ka^\phi_k,\;\;
 N_{3} = b^{\dagger\psi}_{k}b^\psi_{k},\;\; N_4=
b^{\dagger\phi}_{k}b^\phi_{k},
\end{eqnarray}
\begin{eqnarray}
 J_{+} = a^{\dagger\psi}_ka^\phi_k, \;
 J_{-} =a^{\dagger\phi}_ka^\psi_k,\;
 J_{0} =\frac{1}{2}(N_{1} - N_{2}),\;
 \\ \nonumber
   M_{+} = b^{\dagger\psi}_{k}b^\phi_{k},\;
  M_{-} =b^{\dagger\phi}_{k}b^\psi_{k},\;
   M_{0} =\frac{1}{2}(N_{3} - N_{4}),\;
\end{eqnarray}
satisfying the su(2) algebra
\begin{eqnarray}
[J_{+},J_{-}]= -2J_{0},\; [J_{+},J_{0}] =
-J_{+},\;[J_{-},J_{0}]=J_{-},
\end{eqnarray}
\begin{eqnarray}
[M_{+},M_{-}]= -2M_{0},\; [M_{+},M_{0}] =
-M_{+},\;[M_{-},M_{0}]=M_{-}
\end{eqnarray}
and
\begin{eqnarray}
 K_{+} = a^{\dagger\psi}_k b^{\dagger\phi}_{-k},\;
  K_{-} = b^{\phi}_{-k} a^{\psi}_{k},\;
   K_{0} =\frac{1}{2}(N_{1} + N_{3} + 1),\;\nonumber\\
  L_{+} = a^{\dagger\phi}_k b^{\dagger\psi}_{-k},\;
  L_{-} = b^{\psi}_{-k} a^{\phi}_{k},\;
 L_{0} =\frac{1}{2}(N_{2} + N_{4} + 1),\;
\end{eqnarray} satisfying the su(1,1) algebra
\begin{eqnarray}
[K_{+},K_{-}]= 2K_{0},\;
[K_{+},K_{0}]=K_{+},\;[K_{-},K_{0}]=-K_{-},
\end{eqnarray}
\begin{eqnarray}
[L_{+},L_{-}]= 2L_{0},\;
[L_{+},L_{0}]=L_{+},\;[L_{-},L_{0}]=-L_{-}.
\end{eqnarray}
In terms of these generators we can write the Hamiltonian as:
\begin{eqnarray}
H&=&\int d^3k [{2\omega_{\psi}\over a^3}K_0 + {2\omega_{\phi}\over
a^3} L_0 +\frac{\omega_\psi(t)}{a^3}
 \left[\frac{\Omega_\psi^2(t)}{\omega_\psi^2} - 1\right]
(J_{+} + M_{+})
 +\frac{\omega_\phi}{a^3}
\left[\frac{\Omega_\phi^2(t)}{\omega_\phi^2} -
1\right](J_{-} + M_{-}) \nonumber\\ &&+\frac{\omega_\psi}{a^3}
\left[\frac{\Omega_\psi^2(t)}{\omega_\psi^2} - 1\right] ( K_{+} +
L_{-}) +\frac{\omega_\phi}{a^3}
\left[\frac{\Omega_\phi^2(t)}{\omega_\phi^2} - 1\right]
 (L_{+} +  K_{-})],
\end{eqnarray}
The su(1,1) and su(2) symmetries of H are now manifest.We
can digonalize this Hamiltonian by using the following unitary transformation
  \begin{equation}
H'=U^\dagger (R_2)U^\dagger(R_1) H U(R_1)U(R_2),
\end{equation}
where
\begin{equation} U(R_1)U(R_2)=exp[\theta(J_{+}e^{2i\xi}+
J_{-} e^{2i\xi})]exp[\theta(M_{+}e^{2i\xi}+M_{-}e^{2i\xi})],
\end{equation}
and operator $U(R_1)$ provides the well known transformation relations:
\begin{eqnarray}
U^{\dag}(R_1)\left(\begin{array}{c}a^\psi_k
\\a^{\phi}_k\end{array}\right)U(R_1) = \left(\begin{array}{cc}
cos(\theta)& e^{2i\xi}sin(\theta)\\-e^{-2i\xi}sin(\theta)&
cos(\theta)\\

\end {array}\right)
\left(\begin{array}{c} a^\psi_k \\ a^{\phi}_k
\end {array}\right)=
\left(\begin{array} {c} A_k \\ B_k
\end{array}\right),
\end {eqnarray}
while $U(R_2)$ provides the relation
\begin{eqnarray}
U^{\dag}(R_2)\left(\begin{array}{c}b^\phi_{-k}
\\b^{\psi}_{-k}\end{array}\right)U(R_2) = \left(\begin{array}{cc}
cos(\theta)& e^{2i\xi}sin(\theta)\\-e^{-2i\xi}sin(\theta)&
cos(\theta)\\

\end {array}\right)
\left(\begin{array}{c} b^\phi_{k} \\ b^{\psi}_{k}
\end {array}\right)=
\left(\begin{array}{c} C_{k} \\ D_{k}
\end{array} \right),
\end {eqnarray}
and the angle $\theta$ is determined from the relation
$tan(\theta)=\sqrt{\omega_\phi/\omega_{\psi}}$ so that
$cos(\theta)=\sqrt{\omega_\psi/(\omega_{\phi}+\omega_{\psi})}$ and
$sin(\theta)=\sqrt{\omega_\phi/(\omega_{\phi}+\omega_{\psi})}$.
The inflaton field is
 $\eta = |\eta| e^{i\xi}$
and $\eta^* =|\eta| e^{-i\xi}$.

After the rotations\footnote{This gives us the quantum optical
analogy which we are exploiting in this work, as it is analogous
to the effect of a lossless beam splitter. $r=cos(\theta)$ and
$t=sin(\theta)$ are the reflection and transmission coefficients
and the phase shift $\xi$ is between the reflected and transmitted
fields.}  the Hamiltonian assumes the form:
\begin{eqnarray}
H'=U^{\dag}(R_2)U^{\dag}(R_1)HU(R_1)U(R_2) &=&\int d^3k
\frac{(\omega_{\phi}+\omega_{\psi})}{a^3}[\beta(t)[A^\dagger_k A_k +
C_{-k} C^\dagger_{-k} ]\nonumber\\&&+ \alpha(t)[ B^\dagger_{-k} B_{-k} +  D_{k} D^\dagger_{k}]
\nonumber\\ &&+\frac{\omega_\psi}{a^3}
 \left[\frac{\Omega_\psi^2(t)}{\omega_\psi^2} - 1\right][A^\dagger_kC^\dagger_{-k}
+D_{k}B_{-k}]\nonumber\\&&+\frac{\omega_\phi}{a^3}
 \left[\frac{\Omega_\phi^2(t)}{\omega_\phi^2} - 1\right]
[B^\dagger_{-k}D^\dagger_{k}+C_{-k}A_k]].
\end{eqnarray}

This Hamiltonian has a su(1,1)$\times$ su(1,1) dynamical symmetry seen by defining:
\begin{eqnarray}
D_{1+}=A^\dagger_kC^\dagger_{-k},\; D_{1-}=C_{-k}A_k,\;
D_{10}=\frac{1}{2}(A^\dagger_k A_k +C^\dagger_{-k} C_{-k} + 1), \\
D_{2+}=B^\dagger_{-k}D^\dagger_{k},\; D_{2-}=D_{k}B_{-k}\;
D_{20}=\frac{1}{2}(B^\dagger_{-k} B_{-k} + D^\dagger_{k} D_{k} +1),
\end{eqnarray}
satisfying the su(1,1) algebras
\begin{eqnarray}
[D_{1+},D_{1-}]= 2D_{10},\;
[D_{1+},D_{10}]=D_{1+},\;[D_{1-},D_{10}]=-D_{1-},
\end{eqnarray}\begin{eqnarray}
 [ D_{2+},D_{2-}]= 2D_{20},\;[ D_{2+},D_{20}]= D_{2+},\;
[D_{2-},D_{20}]= -D_{2-}.
\end{eqnarray}

Once again this symmetry is manifest through rewriting the Hamiltonian as
\begin{eqnarray}
H'&=&\int d^3k \frac{2(\omega_{\phi}+\omega_{\psi})}{a^3}[[\beta(t)
D_{10}  + \alpha(t) D_{20}]\nonumber\\ &&
+\frac{\omega_\psi}{a^3}
 \left[\frac{\Omega_\psi^2(t)}{\omega_\psi^2} - 1\right][D_{1+}
+D_{2-}]+\frac{\omega_\phi}{a^3}
 \left[\frac{\Omega_\phi^2(t)}{\omega_\phi^2} - 1\right]
(D_{2+}+D_{1-})],
\end{eqnarray}
where:
\begin{eqnarray}
2 \alpha &=&1 - {\lambda\vert\eta\vert^2 a^6\over
2\omega_\phi\omega_\psi},\\
2\beta &=&1 + {\lambda \vert\eta\vert^2 a^6\over
2\omega_\phi\omega_\psi}.
\end{eqnarray}
Since an su(1,1) symmetry implies diagonalizability through
a squeezing (Bogolubov) transformation from the quantum optical
analogies that we are applying, we see here that a product of two
squeezing transformations will provide us with a diagonal Hamiltonian.
The squeezing transformations we use are given by the following
product of squeezing operators
\begin{equation} S(\zeta_1)S(\zeta_2)=exp[\zeta_1D_{1+}
-\zeta_{1}^* D_{1-}]exp[\zeta_2 D_{2+} - \zeta^*_2D_{2-}],
\end{equation}
where, $\zeta_1=r_1exp[i\gamma_1]$ and $\zeta_2=r_2exp[i\gamma_2]$
are the squeezing parameters.\\ The operators $S(\zeta_1)$and $
S(\zeta_2)$ provide the relevant Bogolubov transformations for the
Hamiltonian given in equation (29) in terms of the creation and
annihilation operators $A_k$, $B_k$, $C_k$ and $D_k$ etc.
\begin{eqnarray}
A_s(k,t) = \mu_1 A_k + \nu_1 C^\dagger_{-k},\\
A_s^\dagger(k,t) = \mu_1^* A_k^\dagger + \nu_1^* C_{-k},\\
B_s(k,t) = \mu_2 B_{-k} + \nu_2 D^\dagger_{k},\\
B_s^\dagger(k,t) = \mu_2^* B_{-k}^\dagger + \nu_2^* D_{k},
\end{eqnarray}
where $\mu_1=cosh(r_1)={2\alpha(t)\over(4\alpha(t)-1)^{1\over
2}}$,
$\nu_1=e^{-i\gamma_1}sinh(r_1)=e^{-i\gamma_1}{2\alpha(t)-1\over(4\alpha(t)-1)^{1\over
2}}$, $\mu_2=cosh(r_2)={2\beta(t)\over (4\beta(t)-1)^{1\over 2}}$
and
$\nu_2=e^{-i\gamma_2}sinh(r_2)=e^{-i\gamma_2}{2\beta(t)-1\over(4\beta(t)-1)^{1\over
2}}$ .\\ It is to be noted that obvious mode(k) dependence has
been suppressed and will be made explicit when required. Thus the
final diagonalized Hamiltonian is
\begin{equation}
H_f=\int d^3k\frac{(\omega_{\phi}+\omega_{\psi})}{ 2 a^3}
(4\alpha(t)-1)^{1\over 2}[A^\dagger_s(k,t)A_s(k,t)+1] +\frac{{(\omega_{\phi}+\omega_{\psi})}}{
2 a^3 }(4\beta(t)-1)^{1\over 2}[B^\dagger_s(k,t)B_s(k,t)+1].
\end{equation}
Let $|0(t), 0(t)>$ be the vacuum state of $H_f(t)$. Then, the su(1,1)
symmetry implies that its evolution is generated by the product of squeezing
operators $S(\zeta_1)$ $S(\zeta_2)$ given above. We can write this in terms
of the generators of the su(1,1) $\times$ su(1,1) symmetry as:
\begin{eqnarray}
\vert0(t),0(t)> = e^{\int\frac{d^3k}{(2\pi)^3}\zeta_1(D_{1+}-D_{1-})\zeta_2(D_{2+}-D_{2-})}\vert0(k),0(k)> \label{timeevol}
\end{eqnarray}
where  $|0(k),0(k)> $, vacuum state for H is defined as
$a^\phi_k|0(k),0(k)> = 0$, $a^\psi_k|0(k),0(k)> = 0$,
$b^\phi_{-k}|0(k),0(k)> = 0$, and $b^\psi_{-k}|0(k),0(k)> = 0$.

For a given Lie algebra, a Baker-Campbell-Hausdorff formula can be used to simplify the exponentials of operators in (\ref{timeevol}) and is called disentangling in quantum optics \cite{Perelomov}.
The disentangling formula for su(1,1) is
 \begin{eqnarray}
 e^{\zeta_1(D_{1+}-D_{1-})} e^{\zeta_1(D_{1+}-D_{1-})} \vert0(k),0(k)> =
e^{\gamma_{1k} D_{1+}}e^{\eta'_1 D_{10}}e^{\gamma'_{1k} D_{1-}}e^{\gamma_{2k} D_{2+}}e^{\eta'_2 D_{20}}e^{\gamma'_{2k} D_{2-}}\vert0(k),0(k)>
\end{eqnarray}
where, $\gamma_{1k}=tanh(\zeta_{1k})$ and $\eta'=2ln(cosh(\zeta_{1k}))=-ln(1-\vert\gamma_{1k}\vert^2)$,
 $\gamma'_{1k}=\gamma^*_{1k}$, $\gamma_{2k}=tanh(\zeta_{2k})$ and $\eta'=2ln(cosh(\zeta_{2k}))=-ln(1-\vert\gamma_{2k}\vert^2)$,
 $\gamma'_{2k}=\gamma^*_{2k}$
 and the modes are independent. This allows us to express the
 time evolution of the state as follows \cite{Perelomov}
 \begin{eqnarray}
|0(t),0(t)>& =& \frac{1}{cosh(\zeta_1(t))cosh(\zeta_2(t))}\nonumber\\
&&\prod_k exp(tanh(\zeta_1(t))D_{1+})exp(tanh(\zeta_2(t))D_{2+})|0(k),0(k)> .
\end{eqnarray}
This is just a generalized su(1,1) squeezed state\cite{qo}.\\
QFT in curved space times has taught us the importance
of Bogolubov transformations and consequent population of
one vacuum with particles of another vacuum at an earlier time.
Here, we see this phenomenon with respect to both the squeezing
(su(1,1)) symmetries.
We see that the vacuum at times(t) is populated with particles and anti-particles
with respect to vacuum at t=0.\\
The number of particles and anti-particles can be calculated by the relationship
between the creation and annihilation operators of the initial quanta $a_k^\psi$,
$b_k^\psi$, $a_k^\phi$ and $b_k^\phi$, before reheating with the final creation and annihilation
operators $A_s$ and $B_s$ given by
\begin{eqnarray}
A_s(k,t) &=& (\mu_1cos(\theta))a_k^\psi + (\nu_1sin\theta
e^{2i\xi}) b_{-k}^{\dagger\phi} + (\mu_1 sin\theta e^{2i\xi})
a_k^\phi + (\nu_1cos(\theta)) b_{-k}^{\dagger\psi},\\ B_s(k,t) &=&
(\mu_2cos(\theta)) a_{-k}^\psi + (\nu_2sin\theta e^{-2i\xi})
b_{k}^{\dagger\phi} + (\mu_2 sin\theta e^{-2i\xi})  a_{-k}^\phi +
(\nu_2cos(\theta)) b_{k}^{\dagger\psi}.
\end{eqnarray}

Note that it is the $\phi$ field which carries the baryon number.
The number of baryons at time t is then given by
\begin{equation}
N_B(t) =\sum_k\langle A_s^\dagger(k,t) A_s(k,t)\rangle  =\sum_k \vert\nu_{k1}\vert^2,
\end{equation}
while the number of anti-baryons is given by
\begin{equation}
 N_{\overline{B}}(t)= \sum_k\langle B_s^\dagger(k,t) B_s(k,t)\rangle  = \sum_k\vert\nu_{k2}\vert^2.
\end{equation}

Therefore we have a baryon asymmetry which is given by
\begin{eqnarray}
N_B(t)-N_{\overline{B}}(t) =\sum_k(|\nu_{k1}|^2-|\nu_{k2}|^2),\\
 =\sum_k\left[\frac{\lambda a^6\vert\eta\vert^2}{(\omega_\phi\omega_\psi-\lambda\vert\eta\vert^2a^6)}\right]^2
-\left[\frac{\lambda a^6\vert\eta\vert^{2}}{(\omega_\phi\omega_\psi+\lambda\vert\eta\vert^{2}a^6)}\right]^2.
\end{eqnarray}
where the k dependence is through $ \omega_\phi$, $\omega_\psi$\\
For $\lambda \ll1$, i.e the perturbative limit we recover the results of
Papastamatiou and Parker,
the asymmetry being zero to order$\lambda$,
and non zero at order $\lambda^2$ and above.

 \section{Evolution Of Asymmetry Parameter:}

 In order to get some exact results and numerical values for the  parameter
 providing the baryon asymmetry of the universe,
 we consider a (quite realistic) situation where we can evaluate the Bogolubov
 coefficients exactly.\\
 To this end, we consider the time evolution of wave function
  under the action of the Hamiltonian $H_f$ given by (42).\\

 We go over to the coordinate representation, by defining the operators
 \begin{eqnarray}
A_s(k,t) &=&\frac{e^{i\int\frac{\Omega_+(t)}{a^3}dt}}{2\sqrt{\frac{\Omega_+(t)}{a^3}}}
(\frac{\Omega_+(t)}{a^3}\Pi_A(k,t) + iP_{\Pi_A}(k,t)), \\
A_s^\dagger(k,t) &=&\frac{e^{-i\int\frac{\Omega_+(t)}{a^3}dt}}{2\sqrt{\frac{\Omega_+(t)}{a^3}}}
(\frac{\Omega_+(t)}{a^3}\Pi_A (k,t)- iP_{\Pi_A}(k,t)), \\
B_s (k,t)&=&\frac{e^{i\int\frac{\Omega_-(t)}{a^3}dt}}{2\sqrt{\frac{\Omega_-(t)}{a^3}}}
(\frac{\Omega_-(t)}{a^3}\Pi_B (k,t)+iP_{\Pi_B}(k,t) ), \\
B_s^\dagger (k,t)&=&\frac{e^{-i\int\frac{\Omega_-(t)}{a^3}dt}}{2\sqrt{\frac{\Omega_-(t)}{a^3}}}
(\frac{\Omega_-(t)}{a^3}\Pi_A (k,t)- iP_{\Pi_B}(k,t)).
\end{eqnarray}

$H_{f}$ then reduces to the following:
\begin{equation}
H_f(t)=\int\frac{d^3k}{(2\pi)^3}\frac{1}{2}\left[(\frac{\Omega_+}{a^6})^2\Pi^2_A(k,t)+P_{\Pi_A}(k,t)+
(\frac{\Omega_-}{a^6})^2\Pi^2_B(k,t)+P_{\Pi_B}(k,t)\right],
\end{equation}
where
\begin{eqnarray}
(\frac{\Omega_+}{a^6})^2 =\frac{(\omega_{\phi}+\omega_{\psi})^2}{ 4 a^6}
(4\alpha(t)-1),\\
(\frac{\Omega_-}{a^6})^2 =\frac{(\omega_{\phi}+\omega_{\psi})^2}{
4 a^6 }(4\beta(t)-1).
\end{eqnarray}

The time evolution of a wave function $\chi(t)$ under the action of a Hamiltonian H(t) is simply
\begin{equation}
H(t)\chi(t)=i\frac{d}{dt}\chi(t),
\end{equation}
From the form of $H_f$ given above, it is clear that it is the
direct sum of two independent Hamiltonian $ H_{A}(t)$ and $ H_{B}(t)$
for each of the $A_s$ and $B_s$ modes. Therefore, the wave function
 $\chi(t)$ for the Hamiltonian $H_f$ is just the sum of two wavefunctions
 $ \chi_{A}(t)$ and $\chi_{B}(t)$ where they evolve independently.\\
 Fourier decomposing the wave function  $ \chi_{A}(t)$ and $\chi_{B}(t)$
  and the Hamiltonian $ H_{A}(t)$ and $ H_{B}(t)$, we write

 \begin{eqnarray}
\chi_{A}(t) = \int\frac{d^3k}{(2\pi)^3}\chi_{A}(k,t);\;
H_{A}(t) = \int\frac{d^3k}{(2\pi)^3}H_{A}(k,t)\\
\chi_{B}(t) = \int\frac{d^3k}{(2\pi)^3}\chi_{B}(k,t);\;
H_{B}(t) = \int\frac{d^3k}{(2\pi)^3}H_{B}(k,t),
\end{eqnarray}
 and each mode evolves as.
 \begin{eqnarray}
H_{A}(k,t)\chi_{A}(k,t)=i\frac{d}{dt}\chi_{A}(k,t),\\
H_{B}(k,t)\chi_{B}(k,t)=i\frac{d}{dt}\chi_{B}(k,t).
\end{eqnarray}
Since the Hamiltonians $H_{A}$ and $H_{B}$ are time dependent
harmonic oscillators, in the coordinate space representation
($\Pi_A, P_A$) and ($\Pi_B, P_B$) the wave functions
$\chi_{A}(k,t)$ and $\chi_{B}(k,t)$ can be represented by
gaussian.wave functions.
\begin{eqnarray}
\chi_{A}(k,t) = L_{A}(t)e^{[-W_i(t)\Pi_{A}^2]}\chi_{A}(k,0),\\
\chi_{B}(k,t) = L_{B}(t)e^{[-W_i(t)\Pi_{B}^2]}\chi_{B}(k,0),
\end{eqnarray}
A time derivative gives
\begin{equation}
i\frac{\partial}{\partial t}\chi_{A}(k,t) = \left(i\frac{\dot{L}_{A}}{L_{A}} -
i\Pi^2_{A}\dot{W_{A}}(k,t)\right)\chi_{A}(k,t),
\end{equation}
similarly for $\chi_{B}(k,t)$\\
while the evolution equation (62) gives
\begin{equation}
i\frac{\partial}{\partial t}\chi_{A}(k,t) = \frac{1}{2a^6}\left[\Omega^2_{A}\Pi^2_{A} - \frac{\partial^2}{\partial\Pi_{A}^2}\right]\chi_{A}(k,t),
\end{equation}
A similar equation holds for $\chi_{B}(k,t)$.\\
This gives us
\begin{eqnarray}
W_{A}(t) =-ia^3/2 \frac{\dot{\chi}_{A}(k,t)}{\chi_{A}(k,t)},\\
\end{eqnarray}
implying
\begin{eqnarray}
\ddot{\chi}_{A}(k,t) + 3\frac{\dot{a}}{a}\dot{\chi}_{A}(k,t) + \frac{\Omega^2_{\Pi_{A}}}{a^6}\chi_{A}(k,t) = 0.
\end{eqnarray}

The equations satisfied by the two wave functions for each mode are
\begin{eqnarray}
\ddot{\chi}_{A}(k,t) + 3\frac{\dot{a}}{a}\dot{\chi}_{A}(k,t)+\frac{\Omega^2_+}{a^6}\chi_{A}(k,t) &=& 0,\\
\ddot{\chi}_{B}(k,t) + 3\frac{\dot{a}}{a}\dot{\chi}_{B}(k,t)+\frac{\Omega^2_-}{a^6}\chi_{B}(k,t) &=& 0.
\end{eqnarray}
Our aim in this paper has been to calculate the
baryon asymmetry of the universe for the specific
model we have chosen. To this end, we have now
obtained the time evolution equations for the modes of
the wave functions of our baryons and anti-baryons.
To solve these equations, we shall choose the universe
to be FRW as mentioned earlier.\\
The metric of our FRW is chosen to be conformally flat.
\begin{equation}
ds^2 =a(\tau)^2(d\tau^2 - dx^2).
\end{equation}
by employing the scaled time
\begin{equation}
d\tau = a(t)^{-1}dt.
\end{equation}
The equations of motion for the wave functions
$\chi_{A}(k,t)$ and $\chi_{B}(k,t)$ given above
can be transformed into ones that resemble a
harmonic oscillator with time dependent frequencies.
\begin{equation}
\frac{1}{a^2(\tau)}\frac{d^2\chi_{A}(k,\tau)}{d\tau}+\frac{2}{a^3(\tau)}\frac{da}{d\tau}\frac{d\chi_{A}(k,\tau)}{d\tau}
+\frac{\Omega_-}{a^6}\chi_{A}(k,\tau)=0,
\end{equation}
\begin{equation}
\frac{1}{a^2(\tau)}\frac{d^2\chi_{B}(k,\tau)}{d\tau}+\frac{2}{a^3(\tau)}\frac{da}{d\tau}\frac{d\chi_{B}(k,\tau)}{d\tau}
+\frac{\Omega_+}{a^6}\chi_{B}(k,\tau)=0.
\end{equation}

A change of variable
\begin{equation}
u_{A} = a\chi_{A}
\end{equation}
and similarly for $\chi_B(k,\tau)$ transforms (75), (76) to
\begin{equation}
-u_{A}'' + (E - V_1(\eta\tau))u_{A} =0,
\end{equation}
and
\begin{equation}
-u_{B}'' +(E +  V_2(\eta\tau))u_{B} =0,
\end{equation}
where prime denotes differential with respect to $\tau$ and
\begin{eqnarray}
E = -\frac{(\omega_{\phi}+\omega_{\psi})^2}{4a^6}
\end{eqnarray}
\begin{equation}
V_1(\eta\tau) = -\frac{1}{a(\tau)}\frac{d^2a}{d\tau^2} - [A\vert\eta\vert^2],
\end{equation}
\begin{equation}
V_2(\eta\tau) = \frac{1}{a(\tau)}\frac{d^2a}{d\tau^2} - [A\vert\eta\vert^2].
\end{equation}

where $A =\frac{\lambda(\omega_{\phi}+\omega_{\psi})^2}{8\omega_\phi\omega_\psi} $, and  $\omega_\phi$, $\omega_\psi$ are given by the equation (12)
and (13).\\

These forms of the equations suggest interpretations in two ways.
Treating $\tau$ as a spatial variable, they are Schroedinger
like equations with E=$\omega_k^2(\tau)$. This
allows the calculation of reflection and transmission
coefficients over potential barriers provided by $V_1(\eta \tau)$
and  $ V_2(\eta \tau)$ terms. On the other hand,
they are equations for time dependent harmonic oscillators with time
dependent frequencies (parametric oscillators).\\

In particle creation problems, the potential barrier reflection
($R$) and transmission ($T$) coefficients can be related to the
squeezing parameter ($r$) through $sinh^2(r)=|\nu|^2=\frac{R}{T}$.
We shall use this relationship for the calculation of the
squeezing-parameter-dependent number operator N(k) =$|\nu|^2$
involved in the evolution of wave functions for the particles and
anti-particles in our FRW background.\\ Following previous work
\cite{brand} we assume an oscillating inflaton background
\begin{equation}
\eta  = \Lambda sin(m\tau),
\end{equation}
where $\Lambda=|\Lambda|e^{i\xi}$ is  complex  and m is the
inflaton mass. Substituting the value of $\eta$ in eq(78) and (79)
we get the following differential equations,

\begin{equation}
u_{A}'' + (-\frac{a''}{a} + E - \frac{B}{2} + \frac{B}{2}cos(2m\tau))u_{A} =0,
\end{equation}
and
\begin{equation}
u_{A}'' + (-\frac{a''}{a} + E +\frac{B}{2} -\frac{B}{2} cos(2m\tau))u_{A} =0,
\end{equation}
where $B=|\Lambda|^2 A,$

These are exact equations for the evolution of the baryon and
anti-baryon fields in an expanding universe. Since the number of
baryons and anti-baryons are related to the reflection and
transmission coefficients we see from the above that there will be
an asymmetry, since the baryons encounter a potential barrier
whereas the anti-baryons encounter a potential well.\\ To analyze
these equations for the baryon asymmetry, we first consider the case
of constant expansion ($\frac{a''}{a}=0$).

We see then that equations (84) and (85) are Mathieu equations
\begin{equation}
u_{A}'' + ( E - \frac{B}{2} + \frac{B}{2}cos(2m\tau))u_{A} =0,
\end{equation}
and
\begin{equation}
u_{A}'' + ( E +\frac{B}{2} -\frac{B}{2} cos(2m\tau))u_{A} =0,
\end{equation}

From the theory of Mathieu equations and parametric resonance we see
that the frequency
\begin{eqnarray}
\omega_k^2 =\frac{2\underline{k}^2 +  m_\phi^2 + m_\psi^2 +
2\sqrt{\underline{k}^2 + m_\phi^2 }\sqrt{\underline{k}^2 + m_\phi^2 }}{4}  =(\frac{n}{2} \omega)^2,
\end{eqnarray}
must be half integer multiples of a lowest frequency $\omega$.
The resonance for the lowest frequency occurs for
\begin{eqnarray}
\omega_k^2 - (\frac{n}{2}\omega)^2\equiv \triangle_n,
\end{eqnarray}
where $\triangle_n$ is the width of the the instability band of Mathieu equations.
For the values of $\omega_k$ in this instability band, we assume a broad band
resonance such that the Mathieu equation has instability bands within
which parametric resonance can occur. We shall select the first instability region
as a broad resonance band. The frequency in a broad resonance band is slowly varying,
thus
\begin{eqnarray}
\vert\frac{\omega'_k}{\omega_k}\vert \ll \omega .
\end{eqnarray}
Since $\omega\propto\frac{k}{a}$ this implies that
\begin{eqnarray}
\frac{a'}{a} \ll \omega.
\end{eqnarray}
Thus $\frac{a''}{a}$ being put to zero is a justifiable approximation.\\
Now we make another approximation.\\
We will assume that $m_\phi^2, m_\psi^2 \ll \omega_k^2$. Thus, to a good approximation,
\begin{eqnarray}
 \omega\propto\underline{k}
\end{eqnarray}
and we find
\begin{equation}
u_{A}'' + ( E - B sin^2(m\tau))u_{A} =0,
\end{equation}
and
\begin{equation}
u_{B}'' +( E +  B sin^2(m\tau))u_{B} =0,
\end{equation}
where
\begin{eqnarray}
E = \underline{k}^2,\; A=\lambda,\; B=\lambda|\Lambda|^2
\end{eqnarray}
In the region of broad resonance we replace the oscillating potential near its zeros
with an asymptotically flat potential of the form
\begin{eqnarray}
\vert\eta\vert^2 = |\Lambda|^2sin^2(m\tau)\simeq
2|\Lambda|^2tanh^2(m\frac{(\tau-\tau_i)}{\sqrt{2}}).
\end{eqnarray}
In each instability region ,  we have
\begin{equation}
u_{A}'' + ( \underline{k}^2 - 2\lambda|\Lambda|^2
tanh^2(\frac{m(\tau-\tau_i)}{\sqrt{2}}))u_{A} =0,
\end{equation}
and
\begin{equation}
u_{B}'' +( \underline{k}^2 + 2\lambda|\Lambda|^2
tanh^2(\frac{m(\tau-\tau_i)}{\sqrt{2}}))u_{B} =0.
\end{equation}
We will calculate the transmission and reflection coefficients
of the above equations, which will then provide the
amount of particle production for each case.

The differential equations (97) can be solved through the following
definitions
\begin{eqnarray}
  \kappa_{1}^2 = \frac{\underline{k}^2-(2\lambda|\Lambda|^2 )}{\rho^2}, \;
 \rho^2 =\frac{m^2}{2}
\end{eqnarray}
along with a change of variable given by:
\begin{equation}
y=\rho(\tau-\tau_i),
\end{equation}
we get the following differential equation
\begin{equation}
\frac{d^2u_1}{dy^2} + \left[\kappa_{1}^2 +
\frac{(2\lambda|\Lambda|^2)}{\rho^2} sech^2(y)\right]u_1=0.
\end{equation}
The transmission \cite{Flugge} for this barrier is calculable and in terms of physically known values is
\begin{eqnarray}
\vert T\vert^2={sinh^2(\pi\kappa_{1})\over
\left(cos(\pi\sqrt{\frac{2(\lambda|\Lambda|^2 )}{\rho}+\frac{1}{4}})
\right)^2 + sinh^2(\pi\kappa_1)}.
\end{eqnarray}
The amount of particle production $n_{1k}$ is given as
\begin{eqnarray}
n_{1k} = \vert\nu_{1k}\vert^2
=\frac{1-T}{T}=\frac{\left(cos(\pi\sqrt{\frac{2(\lambda|\Lambda|^2
)}{\rho}+\frac{1}{4}}) \right)^2}
{\left(sinh(\pi\kappa_1\right))^2}
\end{eqnarray}

Now for the anti-particle modes the differential equations (98)
governing the evolution of their wave functions can also be solved
exactly. Defining
\begin{equation}
 \kappa_2^2 = \frac{\underline{k}^2 }{\rho^2},
\end{equation}
and a change of variable as before,
\begin{equation}
y = \rho(\tau-\tau_i)
\end{equation}
gives the following reduced differential equation
\begin{equation}
\frac{d^2u_2}{dy^2} + \left[\kappa_2^2 +
\frac{(2\lambda|\Lambda|^2)}{\rho^2}tanh^2(y)\right]u_2=0.
\end{equation}
The amount of particle production for $B_s(k,t)$ modes  can be
calculated from the transmission coefficient and is
given by
\begin{eqnarray}
n_{2k} = \vert\nu_{2k}\vert^2 =
\frac{\left(cosh(\pi\sqrt{\frac{2\lambda|\Lambda|^2}{\rho^2} -
{1\over 4}})\right)^2}
{sinh^2\left(\pi\sqrt{(\frac{2\lambda|\Lambda|^2}{\rho^2}+\kappa_2^2)}\right)
}.
\end{eqnarray}
We  plot the co-wave  number dependent number of baryons and
anti-baryons  in figure 1. and figure 2. for various values of the
parameter $\frac{2\lambda|\Lambda|^2}{m^2}$ .
 As expected, from the differential equations
satisfied by the mode functions, we see that because the baryons
see a potential barrier and the anti-baryons see a potential well,
there is a differential amplification of particle and
anti-particle modes leading to an asymmetry which survives with
time. Therefore, from our model we have generated a baryon
asymmetry in an entirely non-perturbative fashion. The
approximations made are only at the end of derivations and have
been used to illustrate the methodology. It is to be noticed that
the equations (84-85) are exact for an oscillating inflaton field
in an FRW Universe. It is entirely possible to solve these
equations to incorporate the effects of the expansion of the
Universe on the generation of baryon asymmetry. Since we are
relying on parametric resonance for enhancement/suppression of the
particle/anti-particle creation, the restriction to the lowest
instability band of the Mathieu equation is a reasonable one. The
total integrated  baryon asymmetry of the universe is given by
\begin{equation}
N_{B} - N_{\overline{B}} = \int_0^\infty dk k^2 n_{1k} - \int_0^\infty dk k^2 n_{2k}.
\end{equation}

\begin{figure}[htbp]
   \includegraphics[scale=1]{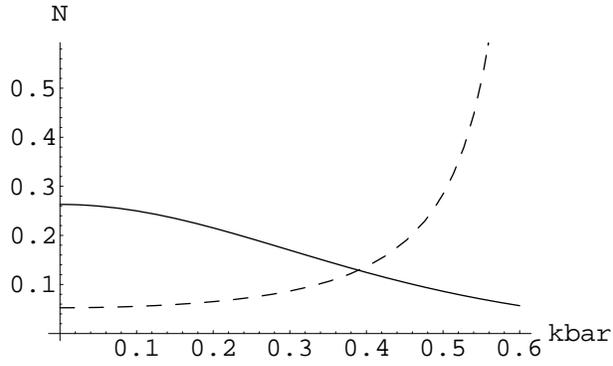}
    \caption{ shows the variation of particles (dashed line) and anti particles
    (solid lines) for  $
    \frac{2\lambda|\Lambda|^2}{m^2}=.5$
    as
    a function of comoving wave number $kbar = \frac{k}{a}$.}
\end{figure}\begin{figure}[htbp]
   \includegraphics[scale=1]{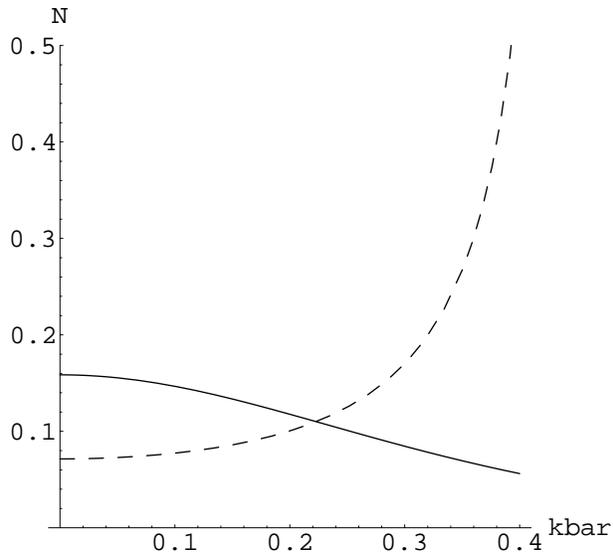}
    \caption{ shows the variation of particles (dashed line) and anti particles
    (solid lines) for  $
    \frac{2\lambda|\Lambda|^2}{m^2}=.25$
    as
    a function of comoving wave number $kbar = \frac{k}{a}$.}
\end{figure}

\section{Conclusions.}
Sakharov's requirements for the generation of a baryon asymmetry in the Universe have been the starting point of our paper.
 Adapting the model of \cite{pp}, to the currently prevalent models of inflation and preheating through a background
  clasical  complex scalar inflaton field \cite{ranga}, we have found equations for the evolution of baryon/anti-baryon modes in an expanding FRW metric.
  We have been able to derive the general evolution equations in an entirely non-perturbative fashion.
   This has been possible through an analogy with quantum optics and its methods. In particular, we have found that at any particular time,
    the Hamiltonian for the quantum modes has $su(2)$ and $su(1,1)$ symmetries.
     Quantum optical techniques allowed us to make these symmetries explicit and thereafter to diagonalize the Hamiltonian.
     The diagonalization procedure further allowed us to relate the rotated and squeezed operators to the traditional Bogolubov transformations between the mode function at an earlier time to those at later times.
The approximations that we have utilized at the end to find explicit variation of the
 number of particles/anti-particles created as a function of the co-moving wave number are only therefore illustrative.
 It is entirely possible to carry out the calculations without these approximations.
  Of course, we will then not have the luxury of the vast literature on the solutions of the Mathieu equations and instability regions allowing for parametric resonant solutions will have to be looked for. On the plus side, the full equations (84-85) allow the possibility of including the contributions to the baryon asymmetry coming from the expansion of the Universe.
 The centerpiece of our effort are the equations [78] and [79],
  which are exact and show the role played
 by the expansion factor and the inflaton field on the evolution of the particle and antiparticle modes.
 These can be solved for any type of inflationary scenario  and
 will be the subject of future investigations.


\begin{thebibliography}{99}
\bibitem{Weinberg} S.~Weinberg, {\it Gravitation and Cosmology} (J Wiley, New York, 1972).
\bibitem{kolb} E.~Kolb and M.~Turner, {\it The Early Universe},
(Addison- Wesley Publishing Company, Redwood City, California,
1990).
\bibitem{sakharov}A.~Sakharov Pis'ma Z. \textit{Eksp. Tero. Fiz}.
  {\bf 5}, 32 (1967);English translation:\textit{JETPLett}. {\bf 5}, 24 (1967).
\bibitem{yoshimura}M.~Yoshimura, \textit{Phys. Rev. Lett.} {\bf 41}, 281 (1978).
\bibitem{weinberg2}D.~V.~Nanopoulous and S.~Weinberg, \textit{Phys. Rev.} {\bf D20}, 2484 (1979).
\bibitem{dolgov}
  A.~D.~Dolgov,
  ``Baryogenesis, 30 years after,''
  arXiv:hep-ph/9707419.
\bibitem{inflation}A.~H.~Guth, \textit{Phys.Rev.} {\bf D23}, 347 (1981).
\bibitem{brand1}
  R.~H.~Brandenberger,
  Braz.\ J.\ Phys.\  {\bf 31}, 131 (2001)
  [arXiv:hep-ph/0102183].
\bibitem{brand}Y.~Shtanov, J.~Traschen and R.~H.~Brandenberger,
Phys.\ Rev.\  {\bf D51}, 5438 (1995).
\bibitem{kofman} L.~Kofman, A.~Linde, A.~A.~Starobinsky, \textit{Phys.Rev.} {\bf D56}, 3258 (1997).
\bibitem{felder} G.~Felder, J.~Garcia-Bellido, P.~B.~Greene, L.~Kofman, A.~Linde and I.~Tkachev,
\textit{Phys. Rev. Lett.} {\bf 87}, 011601 (2001) hep-ph/0012142.
\bibitem{kusenko}
  M.~Dine and A.~Kusenko,
  Rev.\ Mod.\ Phys.\  {\bf 76}, 1 (2004)
  [arXiv:hep-ph/0303065].
\bibitem{garcia}
  J.~Garcia-Bellido,
  ``Tachyonic preheating and spontaneous symmetry breaking,''
  arXiv:hep-ph/0106164.
\bibitem{smit}
  J.~I.~Skullerud, J.~Smit and A.~Tranberg,
  Nucl.\ Phys.\ Proc.\ Suppl.\  {\bf 129}, 771 (2004)
  [arXiv:hep-lat/0309046].
\bibitem{pp}N.~J.~Papastamatiou and L.~Parker, \textit{Phys. Rev.} {\bf D19}, 2283 (1979).
\bibitem{ranga2}R.~Rangarajan,  \textit{Pramana}. {\bf 53}, 1061 (1999).
\bibitem{ranga}R.~Rangarajan, D.~V.~Nanopoulous, \textit{Phys. Rev. }{\bf D65}, 063511 (2001).
\bibitem{japanese}
  K.~Funakubo, A.~Kakuto, S.~Otsuki and F.~Toyoda,
  Prog.\ Theor.\ Phys.\  {\bf 105}, 773 (2001)
  [arXiv:hep-ph/0010266].
\bibitem{brand2}
  K.~R.~S.~Balaji and R.~H.~Brandenberger,
Phys.\ Rev.\ Lett.\  {\bf 94}, 031301 (2005)
  [arXiv:hep-ph/0407090].

\bibitem{cp1}I.~Affleck and M.~Dine \textit{Nucl.phy.} {\bf B249}, 361 (1985).
\bibitem{cp2}
 M.~Riazuddin,
  ``Particle aspects of cosmology and baryogenesis,''
  arXiv:hep-ph/0302020.
\bibitem{bm1}B.~A.~Bambah and C.~Mukku, \textit{Annals of Physics}. {\bf 54}, 314 (2004).
\bibitem{bm2}B.~A.~Bambah and C.~Mukku, \textit{Phys. Rev}. {\bf D70}, 034001 (2004).
\bibitem{bm3}
  B.~Bambah and C.~Mukku,
  ``Charged vs. neutral particle creation in expanding universes: A quantum
  field theoretic treatment,''
  arXiv:hep-th/0307286.
\bibitem{Perelomov} A.~M.~Perelomov, {\it Generalized Coherent States and their Applications},
(Springer- Verlag, Berlin, 1986).
\bibitem{qo}V.V. Dodonov,  \textit{J. Opt. B:Quantum Semiclass. Opt.} {\bf
4} R1-R33(2004).
\bibitem{Flugge} S.~Flugge, \textit{Practical Quantum Mechanics,Vol.I and Vol.II,} (Springer, New York 1976).
\bibitem{Campo} D. Campo and R. Parentani Phys Rev {\bf D67} (2003) 103522
\end{thebibliography}
\end{document}